**Frontier at your fingertips**

Between the nano and micron scales, the collective behaviour of matter can give rise to startling emergent properties that hint at the nexus between biology and physics.

Piers Coleman

*The Hitchhiker's Guide to the Galaxy* famously features a supercomputer, Deep Thought, that after millions of years spent calculating "the answer to the ultimate question of life and the universe", reveals it to be 42. Douglas Adams' cruel parody of reductionism holds a certain sway in physics today. Our 42 is Schroedinger's many-body equation: a set of relations whose complexity balloons so rapidly that we can't trace its full consequences to macroscopic scales. All is well with this equation, provided we want to understand the workings of isolated atoms or molecules up to sizes of about a nanometre. But between the nanometre and the micron, wonderful things start to occur that severely challenge our understanding. Physicists have borrowed the term "emergence" from evolutionary biology to describe these phenomena, which are driven by the collective behaviour of matter.
  Take, for instance, the pressure of a gas – a cooperative property of large numbers of particles that is not anticipated from the behaviour of one particle alone. Although Newton's laws of motion account for it, it wasn't until more than a century after Newton that James Clerk Maxwell developed the statistical description of atoms necessary for an understanding of pressure.
  The potential for quantum matter to develop emergent properties is far more startling. Atoms of niobium and gold, individually similar, combine to form crystals that kept cold, exhibit dramatically different properties. Electrons roam free across gold crystals, forming the conducting fluid that gives it its lustrous metallic properties. Up to about 30 nanometres, there is little difference between gold and niobium. It's beyond this point that the electrons in niobium start binding together into the coupled electrons known as "Cooper pairs". By the time we reach the scale of a micron, these pairs have congregated in their billions to form a single quantum state, transforming the crystal into an entirely new metallic state — that of a superconductor, which conducts without resistance, excludes magnetic fields and has the ability to levitate magnets.
  Superconductivity is only the beginning. In assemblies of softer, organic molecules, a tenth of a micron is already big enough for the emergence of life. Self-sustaining microbes little more than 200 nanometres in size have recently been discovered. While we understand the principles that govern the superconductor, we have not yet grasped those that govern the emergence of life on roughly the same spatial scale.
  In fact, we are quite some distance from this goal, but it is recognized as the far edge of a frontier that will link biology and physics. Condensed matter physicists have taken another cue from evolution, and believe that a key to understanding more complex forms of collective behaviour in matter lies in competition not between species, but between different forms of order. For example, high-temperature superconductors — materials that develop superconductivity at liquid nitrogen temperatures — form in the presence of a competition between insulating magnetic behaviour and conducting metallic behaviour. Multi-ferroic materials, which couple magnetic with electric polarization, are found to develop when magnetism competes with lattice-distorting instabilities**.**
  A related idea is "criticality"— the concept that the root of new order lies at the point of instability between one phase and another. So, at a critical point, the noisy fluctuations of the emergent order engulf a material, transforming it into a state of matter that like a Jackson Pollack painting, is correlated and self-similar on all scales. Classical critical points are driven by thermal noise, but today, we're particularly interested in "quantum phase transitions" involving quantum noise: jigglings that result from Heisenberg's uncertainty principle. Unlike its thermal counterpart, quantum noise leads to diverging correlations that spread out not just in space, but also in time. Even though quantum phase transitions occur at absolute zero, we're finding that critical quantum fluctuations have a profound effect at finite temperatures.

For example, "quantum critical metals" develop a strange, almost linear temperature dependence and a marked predisposition towards developing superconductivity. The space-time aspect of quantum phase transitions gives them a cosmological flavour and there do appear to be many links, physical and mathematical, with current interests in string theory and cosmology. Another fascinating thread here is that like life, these inanimate transformations involve the growth of processes that are correlated and self-sustaining in time.

Some believe that emergence implies an abandonment of reductionism in favour of a more hierarchical structure of science, with disconnected principles developing at each level. Perhaps. But in almost every branch of physics, from string theory to condensed matter physics, we find examples of collective, emergent behavior that share common principles. For example, the mechanism that causes a superconductor to weaken and expel magnetic fields from its interior is also responsible for the weak nuclear force — which plays a central role in making the sun shine. Superconductors exposed general principles that were used to account for the weak nuclear force.

To me, this suggests that emergence does not spell the end for reductionism, but rather indicates that it be realigned to embrace collective behavior as an integral part of our universe. As we unravel nature by breaking it into its basic components, avoiding the problem of "42" means we also need to seek the principles that govern collective behaviour. Those include statistical mechanics and the laws of evolution, certainly, but the new reductionism we need to make the leap into that realm between nano and micron, will surely demand a new set of principles linking these two extremes.

**Piers Coleman is at the Department of Physics and Astronomy, Rutgers University, 136 Frelinghuysen Road, Piscataway, New Jersey 08854-8019, United States.**